# Investigation on QoS of Campus-wide WiFi Networks

N. Sulaiman and C.Y. Yaakub

**Abstract**— WiFi is widely implemented in campus wide including administrative, teaching and student's accommodation. Wireless communications are associated with interconnect devices which includes cellular networks, infrared, bluetooth and WiFi enabled devices. It involves mobility and freedom of assessing information anytime and anywhere. A study on WiFi networks in a campus environment is presented in this paper. The aim of the research was to investigate the connectivity problems to WiFi networks. The study includes WiFi performance analysis as well as network auditing. Channel overlapping and saturation condition were some of the problems encountered. Different types of software were used for analyzing the results.

———————————— ◆ ————————————

## 1 INTRODUCTION

DIFFERENT has different technologies have emerged to enable the a global communication platform for the Internet. Thus, wireless network based products are becoming norms in our daily activities and its application is growing at a rapid pace. Wireless data communication offers many benefits which include better mobility, portability as well as reliability. Wireless technology has both positive as well as negative aspects. The adoption of wireless technology is demanding and it is crucial for a product market success. Wireless product generates group of interest among peer groups which will result in well received product and accepted by wider group of users [1].

The wireless connectivity is used for the exchange of information over the networks and it is also widely used for IP telephony and IP video conferencing. To help the industry address the potential of wireless technology, the IEEE provides standards for its operation. The network IEEE 802.11b standard has been used several years in the communication industries. Table I shows three fundamental standards for IEEE 802.1.

Table I. The 802.11 Standards

| Standard | Radio Bank | Modulation | Max Link Coverage | Max Data Rate |
|---|---|---|---|---|
| 802.11b | 2.4 GHz | DSS | 100m/ 328ft | 11 Mbps |
| 802.11a | 5 GHz | OFDM | 50m/ 164ft | 54 Mbps |
| 802.11g | 2.4 GHz | OFDM | 100m/ 328ft | 54 Mbps |

Since WiFi is a wireless standard, the frequency range offered is limited and governed by various authorities from International to Local. The general public is allowed to use the ISM frequency range between 2.400 to 2.485 GHz for the 2.4GHz Euro Standard. Many products have been developed for the general public utilizing this frequency range, both backhaul and client access. In addition, the power output from the device and antenna is also controlled [2].

With a wireless environment, communication becomes borderless as people could be connected anytime and anywhere. Problems with wireless networks have been studied in [3],[4],[5],[6],[7]. In [8], a web service for location discovery service with wireless LAN connectivity was discussed. Mobility of the users was managed using management and location based services. Diversity mechanism allows efficient communication over fading channels, however analyze and design it in networks with many nodes is difficult [9].

Ensuring a high quality service in wireless network environment is crucial. Accurate tracking and location prediction is one of the ways to significantly improve the performance and reliability of wireless networks protocol and infrastructure. In order to understand the wireless users and to develop a suitable wireless application which will be acceptable to the users, it is important to see the needs of people today and towards the future. Mobility management involves how mobility affects the overall system performance. The current and future wireless mobile networks effectively delivering services to mobile users need to be studied[10].

In [11], the use of portable handheld computers with wireless Internet access to improve teaching and learning was studied. Real time learning in the wireless classroom was discussed. The study involved both local and wide area network environments. Wireless network coverage is one of the fundamental problems in wireless network which was discussed in [12]. The quality of service that can be provided by a particular network was studied.

The instructional redesign of the traditional higher education classroom was studied in [13]. It focused on the environment for anytime, anywhere learning and access. The wireless environment and the use of a real-

————————————————

- *Norrozila Sulaiman is with theUniversiti Malaysia Pahang, Malaysia.*
- *Che Yahaya Yaakub is with the Universiti Malaysia Pahang, Malaysia.*

13time assessment in the classroom using a wireless infrastructure were explained. Similarly, a mixed infrastructure network for data traffic in a home environment was studied in [14] and the aim was to maximise the overall network capacity.

In [15], authors propose methods to integrate ad-hoc operations into the infrastructure mode by using Ad-Hoc Awareness Direct Connection (AHADC) and Direct Cut-Through Forwarding DCTF) schemes. A study on people location system based on WiFi signal measure was presented in [16]. The current location systems based on Wi-Fi were mainly applied in the location of indoor robots using the measure of their communication interface and additional sensors.

## 2 WIRELESS NETWORKS

### 2.1 WiFi Connectivity

Wireless Access Points (AP) were deployed widely in the campus area and poor connectivity to WiFi network at certain areas is one of the major problems. Users have either a Personal Computer or Laptop, and fast connection to the campus real time system as well as to the Internet are their expectations. The cause of the poor Internet connection and the health level of the Access Point were part of the objectives of the investigation.

The WiFi system was implemented a few years back with the intention of offering the users easy connectivity. Nevertheless, it was found that students were having problems in connecting to the WiFi networks mainly at the students accommodation area. Throughout the duration of the project, a few areas were investigated. Fig. 1 shows the campus general layout.

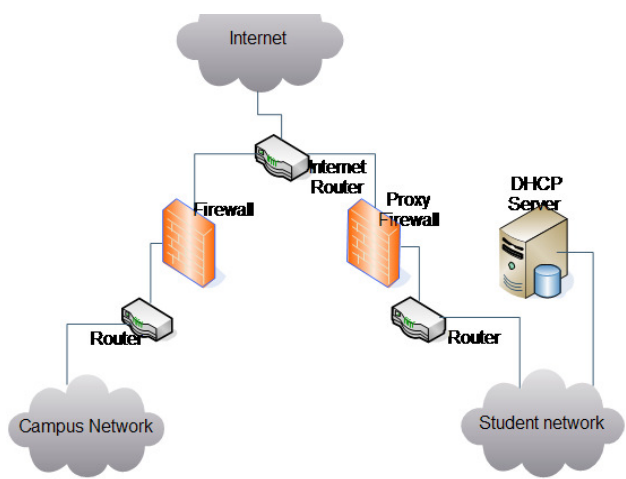

Fig 1. Campus General Network

Student accommodation consists of a few blocks, which contain five floors for each block. There are six wireless devices attached to each block. The device consists of two wireless radios, for every floor and outdoor coverage. The wired network has been redesigned where the network from the student accommodation had been placed into one Virtual Local Area Network (VLAN). The students WiFi networks were also placed in this environment. A separate proxy server and firewall are used to manage the network to the Internet. A router is located behind the firewall, which allows connectivity from the student network to the original campus network. The students are allowed to access the student server without going through the various firewalls as shown in the Figure.

From a core switch, the network to the student accommodation is split into two. A fiber connection is installed from the core switch to a switch at two different blocks. From these two blocks, the fiber connectivities to the other student blocks are established.

The WiFi APs are configured to the Euro standard which offers a frequency range with 13 channels. Each of the AP has a maximum configured capacity of 5.5 Mbps. Each of the AP has its own Service Set Identifier (SSID). For each block there are 6 wireless devices. The study was performed due to analyse the network performance in relations to the usage of the WiFi network. Most of the problems are concerning poor Internet access

In relations to the WiFi coverage, sometimes users were not be able to detect a WiFi node. Although some of the users were able to detect the node, the signal strength was very low. At the same time some users managed to detect very good signal but they can neither access the Internet nor the local server. There was a suggestion that the slow Internet Access could be due to users playing online games within the wireless network.

## 3 TROUBLESHOOTING

To achieve the intended objectives, a study on the WiFi coverage was carried out which involved auditing the networks, both wired and wireless. Saturation level of the Access Point was also investigated.

Different brand of laptops (including clone computers) were selected for the study such as:
1. HP,
2. Asus,
3. Fujitsu,
4. Compaq

Several types of software were used:
1. Fluke Network Inspector,
2. Network Stumbler,
3. Wildpackets AiroPeek
4. NX, Wildpackets iNetTools,
5. WiFi Manager,
6. Commview for WiFi and LviewPro.

### 3.1 WiFi Analysis

A site survey on the student accommodation was carried out and the WiFi Survey was conducted using Network Stumbler which was tested at various locations within each block. WiFi Manager was connected to the wired environment and it was used to measure network performance.

The software was allowed to run over the period of the study, to collect the performance and information concerning utilization of the various WiFi AP or nodes. Furthermore, the application is able to scan the network for IP devices based on Simple Network Management Protocol (SNMP). Thus devices which do not activate the SNMP will not be able to be monitored effectively. The devices which are located out of the current subnet might not be automatically detected.

The software is also capable to detect the health level of applications such as Dynamic Host Configuration Protocol (DHCP) and DNS servers. Moreover, Wildpackets AiroPeek NX was used to analyse the WiFi signal and data packet reliability. If there is any signal clasing or overlapping, the software will be able to show the final strength received at each client WiFi card. The quality of data packets over the WiFi radio can also be analysed. Commview for WiFi was used to verify the connection between Access Points and the associated client.

### 3.2 Network Auditing

Network auditing was conducted to access the reliability of the network both wired and wireless by using Fluke Network Inspector version 5, connected over the wired environment. The application was allowed to run over the period of the study so as to allow sufficient information to be collected. The application is able to scan the network for IP devices based on the SNMP protocol. Thus devices not having the SNMP protocol being activated will not be able to be monitored effectively. Also, devices outside of the current subnet might not be automatically detected. Besides being able to scan for IP devices, the application is able to obtain information concerning the physical ports of the Access Points, both Ethernet and Radio, through SNMP probing. The most important part of the network audit, is the analyses and interpretation of the results obtained

## 4 RESULTS AND DISCUSSIONS

### 4.1 WiFi Analysis

According to [17], 225 reading locations were selected to analyse the WiFi signal detection from various WiFi APs deployed within the accommodation area. At certain locations more than 30 Access Point signals were detected. It was found that many Access Points utilizing the same channel. The signal strength of these Access Points is low and the wireless signal strength was about the same. A cancellation effect would ultimately give rise to a lower effective signal.

As a result, the quality of the connectivity was very poor although the Access Points were detected by the client computer. Moreover, the client would be cutoff very regularly. Overlapping functional frequencies between various Access Points has also result in unreliable data transfer and a higher level of data corruption.

Both, channel clashing and frequency overlapping will cause a high data error and noise within the radio signal. The quality and reliability of the connectivity is greatly threatened. Therefore, with a proper channel selection, problems of data corruption, regular disruption or low connection quality can be reduced. As a result, the reliability of the WiFi coverage can be assured.

Each Access Point is configured with its own SSID and it has a VLAN of its own. Hence, changing the SSID could cause the IP to be dropped and not released. The implementation of unique SSID for each AP, will not allow for roaming to happen. To allow seamless roaming from one AP to another without disruption and having data integrity, the AP should be configured with the same SSID so that they are in the same wireless VLAN (WVLAN). The difference in Signal-to-Noise ratio between these Access Points is too small to give an effective connectivity. Most users would have to manually connect to Access Point.

During the study period, it was found that the maximum data transfer rate detected over a WiFi radio was 5421 Kbps. This shows that a maximum bandwidth was achieved should the Access Point was configured with a bandwidth of 5.5Mbps. The overall utilization was less than one percent (1%) per day and it would give a total of 4.752Mb of data transferred within the day. The result shows that this is a very small amount of data being transferred. The reason could be either low data transferred or the users were not able to get connected. In addition, the number of error packets per second is rather high in some Access Points.

It was found that some AP has a Mean Time to Recovery (MTTR) of more than 12 hours. The Mean Time before Failure (MTBF) in all the Access Points is less than 24 hours. A high MTTR could be caused by interference between the various APs. Overall, it was found that the effective signal strength was low even though the client PC showed high signal strength and part of the reason was due to the placement of the antenna [17].

### 4.2 Network Auditing

In network audit, the physical ports of the APs were monitored for utilization, broadcast and errors during the study period. From the estimated 80 installed units, a total of 44 units were actually detected as APs and 10 were detected only as network devices due to insufficient SNMP information from these units. The rest were not detected either on the Ethernet LAN or as an AP. This was due to either the APs are faulty or network switches are faulty, or both problems.

Ports of the APs represent the physical radio ports. These ports provide connectivity between the radio and the Ethernet port of the APs. All these ports were configured as 11Mbps. A number of these physical radio ports experienced high utilization at above 50% and the highest was 97.8%. This high utilization was seen as a spike and not constantly observed. User data was observed to be low but broadcast traffic was high, which constitutes to the high utilization of the ports at times. The broadcast traffic was mainly management traffic and this includes User Datagram Packet(UDP) and Address Resolution

Protocol(ARP) packets.

The high utilization was also constituted from the high error of data packets at the APs. This was due to the interference of radio channel as explained earlier. All APs reported Incorrect Subnet Mask which is one of the reasons for the urss having difficulty to login as they belong to another Subnet Mask. The Ethernet LAN port is configured for 100Mbps and port utilization was observed to be very low, less than 10% data traffic. There were no observed physical port errors on the Ethernet LAN or the physical radio ports of the AP. These errors should not be seen at all as any error will lead to data packet error that is detrimental to data transmission. Some APs are configured with DHCP capability. This could affect some users who are connected to these APs.

### 4.3 General Discussion

From the results obtained, it was found that too many APs were having their signals clashing and overlapping with each other. Clashing of signals has caused a poor connectivity to a particular AP. At the same time, a high number of retries were seen and these signals clashing were causing the radio portion of the Access Points to fail regularly. The overlapping of signals caused a high level of Cyclic Redundancy Check (CRC) Errors so as the placement of the antenna that has caused some echo effect.

The health of the AP is greatly reduced with a MTTR for some of the Access Points being more than 12 hours. In addition, the MTBF for some of the APs is less than 12 hours, meaning that the AP is expected to fail more than twice daily. A total of at least fifteen (15) switches were expected to be seen, however only seven) were captured. This was could be due to the fact that some of these devices are having their IP being located within a different subnet or VLAN. Therefore, WiFi Manager was not able to capture the traffic from these devices. To view all the devices within a specific area effectively, all devices within the VLAN should have IPs within the defined subnet. This was also reported by Fluke Network Inspector. The WiFi Manager showed that, the DHCP Server that were used to serve IP request to a certain extend was affected by the heavy background management traffic and other ARP and UDP traffic.

The high management traffic together with the high ARP and UDP traffic have loaded the switches within the network to an extend that these devices are showing symptoms of failure. WiFi Manager showed that the ports of the switches are starting to fail so as the devices. Among the reasons for the high UDP traffic could be due to viruses and worms that are looking for vulnerable systems to further infect. It kept the system active for 20 hours before shutting down the system.

The user was not able to use the normal system shutdown or reset. The issues was highlighted as wireless problems, are to a certain extend it was more than just a wireless problem. Although there are issues with the WiFi network, resolving the WiFi network would lift-up part of the problems faced by the users. Optimizing the current WiFi Network with the currently installed products could solve part of the wireless problems but would leave the WiFi coverage with blind spots. The optimal condition is to install the APs outdoors with the antenna located not more than certain distance away from the AP.

Based on the results obtained from the study, the number of clashing and overlapping signals needs to be reduced especially at locations where the signal strength of signals from the same channel are about the same strength. Having the antenna placed within the corridor is causing an echo effect to the AP itself. In addition, there are metal shoe racks and metal cabinets that are in the signal pathway of the antenna to the client PC. The antenna will need to be repositioned or changed to one that can be used in such an environment. This would also reduce the CRC Errors caused by the reflecting signals to the AP. The SSID used need to be planned to improve the roaming facility. The channels used for the APs need to be planned based on the maximum bandwidth allowed for a particular AP so that no overlapping of signals would happen.

## 5 CONCLUSION

The results reveal that wireless networks often have performance problems. It explained the user disappointment with the WiFi based networks that have been deployed. Wireless problems could be due to channel/signal clashing and overlapping, a high amount of CRC Errors, poor placement of antenna, APs that are failing very regularly, and switching over from one SSID to another. The study showed that the utilization over the wireless network is low and only a few APs showed a high saturation level which caused by the Game servers.

The poor Internet Access was due to the client PC not being able to get an IP when switching from one SSID to another. The proxy server to a certain extend was one of the cause of timeouts if it is overloaded. In addition, inappropriate use of the firewall was one of the problems causing a poor Internet Access. The policies deployed should be minimal but effective to protect both internal networks, so as to allow the firewall to function effectively. It is good to have a proxy server for caching purposes when the amount of traffic going to the same place is high. In conclusion, further investigation need to be conducted. From the study, a number of recommendations can be made to improve the network access. This involves rectifying the wireless problems and analysing the problems within the wired network that has caused a high level of background traffic in the form of ARP and UDP packets. In addition, the network application servers that seem to function intermittently also need to be studied. Other network and application components should also be studied to further improve the usability of the network.

### ACKNOWLEDGMENT

The authors wish to thank Universiti Malaysia Pahang for the support on this project.

**N.Sulaiman** Norrozila Sulaiman graduated from Sheffield Hallam University with a BSc (Hons) in Computer Studies in 1994. She worked as an assistant network researcher with Employment Service in UK, as part of the sandwich course programme. After graduated, she worked as an Artificial Intelligence researcher and involved in joint collaboration projects between the governments of Malaysia and Japan for about 5 years. In year 2001, she completed her MSc in Information Technology and involved in a research on an application using Wireless Application Protocol (WAP). She obtained her PhD in the area of mobile communication and networks. Currently, she is a senior lecturer at Faculty of Computer System and Software Engineering, University Malaysia Pahang. Her main research interests include communication networks and information security. She is a member of IEEE.

**C.Y.Yaakub** Che Yahaya Yaakub graduated with a BSc (Hons) in Computer Science from National University of Malaysia in 1988. In 1999, he completed his MSc in Computer Science from University Technology Malaysia. He worked as a System Analyst and he was the Head of ICT centre of University College of Engineering & Technology Malaysia. He is a member of IEEE Computer Society and ACM. Currently, he is a senior lecturer and his main research interests are in wireless network and network performance.